\documentclass[twocolumn,prl,aps,floatfix,epsf,psfig,groupaddress,amssymb]{revtex4}
\usepackage{graphicx,subfigure}
\usepackage{color,soul}
\usepackage{epsfig}
\usepackage{mathtools}
\DeclarePairedDelimiter{\evdel}{\langle}{\rangle}
\newcommand{\ev}{\operatorname{}\evdel}
\usepackage{amsmath}
\usepackage{float}
\begin{document}
\title{Effects of long-range disorder and electronic interactions on the 
optical properties of graphene quantum dots}

\author{A. Alt{\i}nta\c{s}}
\author{K. E. \c{C}akmak}
\author{A. D. G\"u\c{c}l\"u}

\affiliation{Department of Physics, Izmir Institute of Technology, IZTECH,
  TR35430, Izmir, Turkey}

\date{\today}

\begin{abstract}
We theoretically investigate the effects of long-range disorder and
electron-electron interactions on the optical properties of hexagonal
armchair graphene quantum dots consisting of up to 10806 atoms. The
numerical calculations are performed using a combination of tight-binding,
mean-field Hubbard and configuration interaction methods.
Imperfections in the graphene quantum dots are modelled as a
long-range random potential landscape, giving rise to electron-hole
puddles.  We show that, when the electron-hole puddles are present,
tight-binding method gives a poor description of the low-energy
absorption spectra compared to meanfield and configuration interaction
calculation results. As the size of the graphene quantum dot is
increased, the universal optical conductivity limit can be observed in
the absorption spectrum. When disorder is present, calculated
absorption spectrum approaches the experimental results for isolated
monolayer of graphene sheet.
\end{abstract}
\maketitle

\section{Introduction}
Graphene, single layer of carbon atoms arranged in a two-dimensional
honeycomb lattice\cite{wallace1947band,novoselov2004electric}, has
attracted enormous research interest due to superior electronic
electrical
conductivity\cite{novoselov2004electric,novoselov2005two,zhang2005experimental,rycerz2007valley},
mechanical
strength\cite{lee2008measurement,dikin2007preparation,mechanicthermal1},
thermal conductivity\cite{mechanicthermal1} and unique optical
properties\cite{optical1,optical2}.  Moreover, the electronic and
optical properties of graphene can be manipulated at the nanoscale in a
desired way by controlling lateral size, shape, type of edge, doping
level and the number of layers in graphene
nanostructures\cite{nanostructures1,nanostructures2,nanostructures3,nanostructures4,
nanostructures5,nanoribbonedge5}.
Among those various nanostructures of graphene,
graphene quantum dots (GQDs)\cite{nanostructures6,GrapheneQuantumDotBook,quantumdot1,quantumdot2,quantumdot3,
quantumdot4,quantumdot4a,
quantumdot6,quantumdot7,magneticzigzag1,devrimexcitonic2,
devrimmagnetic1,opticzigzag1,opticzigzag2,
opticzigzag3,electronicoptic1} offer a
possibility to simultaneously control the electronic, magnetic and
optical functionalities in a single material.

GQDs are classified according to their edge character since the edges
play an important role in determining electronic, optical and magnetic
properties of GQDs\cite{quantumdot1,quantumdot2,quantumdot3,quantumdot4,quantumdot6,
quantumdot7,quantumdot4a,magneticzigzag1,devrimmagnetic1,devrimexcitonic2}. In particular, armchair and
zigzag edges are the most stable edge
structures\cite{nanoribbonedge5,quantumdot6,quantumdot7} while GQDs with
zigzag edges are found to exhibit unusual
magnetic\cite{quantumdot4a,magneticzigzag1,devrimmagnetic1} and
optical\cite{devrimexcitonic2,opticzigzag1,opticzigzag2,opticzigzag3,electronicoptic1}
properties due to the presence of a degenerate band of states at the
Fermi level. On the other hand, armchair edges do not lead to degenerate band of states at the Fermi level, hence, can be used as small model of bulk graphene which does not have edge states\cite{armchair1}.


Properties of graphene nanostructures fabricated and observed upon
substrates\cite{substrate1,substrate2} may become affected by
imperfections due to the environment and become disordered. In
particular, if the disorder has a long-range character, it can lead to
charge localizations as electron-hole
puddles\cite{puddle1,puddle2,puddle3,puddle4}. For instance, magnetic
properties of graphene nanoribbons are found to be strongly dependent
of long-range impurities\cite{ulasmenafdevrim}. In addition,
the role of electron-hole puddles on the formation of Landau levels
in a graphene double quantum dot was investigated experimentally by
K. L. Chiu \textit{et al.}\cite{disorderpuddle1}.

On the other hand, a striking optical property of graphene is the
universal optical conductivity (UOC) which can be identified as explicit manifestation of light and matter interaction\cite{UOC1,UOC2,opticconductance3}. The experimental observation of UOC for a graphene sheet seems to
indicate that optical properties are robust against imperfections,
although significant deviations from UOC at lower energies was
observed\cite{opticconductance1,opticconductance2}. To our knowledge,
a detailed theoretical investigation of combined effects of long-range
disorder and electron-electron interactions on the optical properties 
of graphene quantum dots is still lacking.

In this work, we investigate theoretically electronic and optical
properties of medium and large sized hexagonal armchair GQDs
consisting of up to 10806 atoms to understand the role of long-ranged
disorder on the optical properties. Our main contribution involves
inclusion of electron-electron interactions within meanfield and
many-body configuration interaction approaches. We show that the 
electron-electron interactions play a significant role in redistributing
electron-hole puddles, thus strongly affecting the optical properties.
We also investigate the large size limit of the GQDs as compared to
optical properties of bulk graphene\cite{opticconductance1,opticconductance2,opticconductance3}
and show that UOC can be observed in GQDs with a diameter of 18 nm.

The paper is organized as follows. In Sec. II, we describe our model
Hamiltonian including electron-electron interaction and random
potential term, and the computational methods that we use in order to
compute optical properties of hexagonal armchair GQDs. The
computational results on the electronic and optical properties are
presented in Sec. III.  Finally, Section IV provides summary and
conclusion.

\section{Method and Model}
In the tight-binding (TB) approach, the one electron states of GQD can
be written as a linear combination of $p_z$ orbitals on every carbon
atom since the $s$, $p_x$ and $p_y$ orbitals are considered to be
mainly responsible for mechanical stability of graphene. Then, within 
the meanfield extended Hubbard approach, Hamiltonian can be written as:

\begin{align}
H_{MFH} =& \sum_{ij\sigma} ( t_{ij} c^{\dagger}_{i\sigma} c_{j\sigma} + h.c) \nonumber \\
&+ U\sum_{i\sigma} (\ev{n_{i\sigma}} - \frac{1}{2})n_{i\bar{\sigma}} + \sum_{ij\sigma} V_{ij} (\ev{n_{j}}-1)n_{i\sigma} 
 \nonumber \\ 
&+\sum_{i\sigma}V_{imp}(i) c^{\dagger}_{i\sigma}c_{i\sigma} 
\end{align}

where the first term represents TB Hamiltonian and $t_{ij}$ are the hopping
parameters given by $t_{nn}=-2.8$ eV for nearest neighbours and
$t_{nnn}=-0.2$ eV for next
nearest-neighbours\cite{hopping1}. The $c^{\dagger}_{i\sigma}$ and
$c_{i\sigma}$ are creation and annihilation operators for an electron
at the $i$th orbital having spin $\sigma$, respectively. Expectation
value of electron densities are represented by
$\ev{n_{i\sigma}}$. The second and third terms represent onsite and
long range Coulomb interaction, respectively. We take onsite
interaction parameter as $U=16.522/ \kappa$ eV and long-range
interaction parameters $V_{ij}=8.64/ \kappa$ and $V_{ij}=5.33/ \kappa$
for the first and second nearest neighbours with effective dielectric
constant $\kappa=6$\cite{dielectricconstant1}, respectively. Distant
neighbor interaction is taken to be $1/d_{ij} \kappa$ and interaction
matrix elements are obtained from numerical calculations by using
Slater $\pi_z$ orbitals \cite{Slatermatrix1}. Last term corresponds to
impurity potential $V_{imp}(i)$ account for substrate effects.

After diagonalizing the mean-field Hubbard (MFH) matrix self consistently
by starting with TB orbitals, we obtain the Hubbard quasi-particle
spectrum which has fully occupied valance band and completely empty
conduction band. Next, in order to take into account two-body configuration interactions (CI), excitonic
correlation effects of electron-hole, we solve the many-body Hamiltonian for a hole and
an electron:

\begin{equation}
\begin{split}
H_{eh}=& \sum\limits_{p'\sigma}^{}\epsilon_{p'}b_{p'\sigma}^{\dagger}b_{p'\sigma} -\sum\limits_{p,\sigma}^{}\epsilon_{p}h_{p\sigma}^{\dagger}h_{p\sigma} \\
&-\sum\limits_{\substack{p',q,r,s' \\ \sigma,\sigma'}}^{}\left\lbrace<rp'\mid V \mid s'q> \right. \\ 
&\left. -\left(1-\delta_{\sigma\sigma'}\right)<rp'\mid V \mid qs'>\right\rbrace
b_{p'\sigma}^{\dagger}h_{q\sigma'}^{\dagger}h_{r\sigma'}b_{s'\sigma} \\
&+\sum\limits_{\substack{p',q,r,s' \\ \sigma,\bar{\sigma}}}^{}<rp'\mid V \mid qs'>b_{p'\bar{\sigma}}^{\dagger}h_{q\sigma}^{\dagger}h_{r\bar{\sigma}}b_{s'\sigma} \\ 
\end{split}
\end{equation}

Here, the first two terms describe electron and hole quasi-particle
energies obtained from the meanfield calculations, third term
describes the electron-hole Coulomb attraction, and the fourth and
fifth terms represent the electron-hole exchange interactions. Indices
with prime denotes electron states and without prime denotes hole
states. The two-body electron-hole scattering matrix elements are
calculated from two-body on-site and long-range Coulomb matrix
elements \cite{GrapheneQuantumDotBook}.

In this work, we consider three different sizes of hexagonal armchair
GQDs (see for example Fig. 1a) consisting of 1014, 5514 and 10806 atoms
and having widths of 5 nm, 13 nm and 18 nm, respectively. 

\begin{figure}
\includegraphics[scale=0.20]{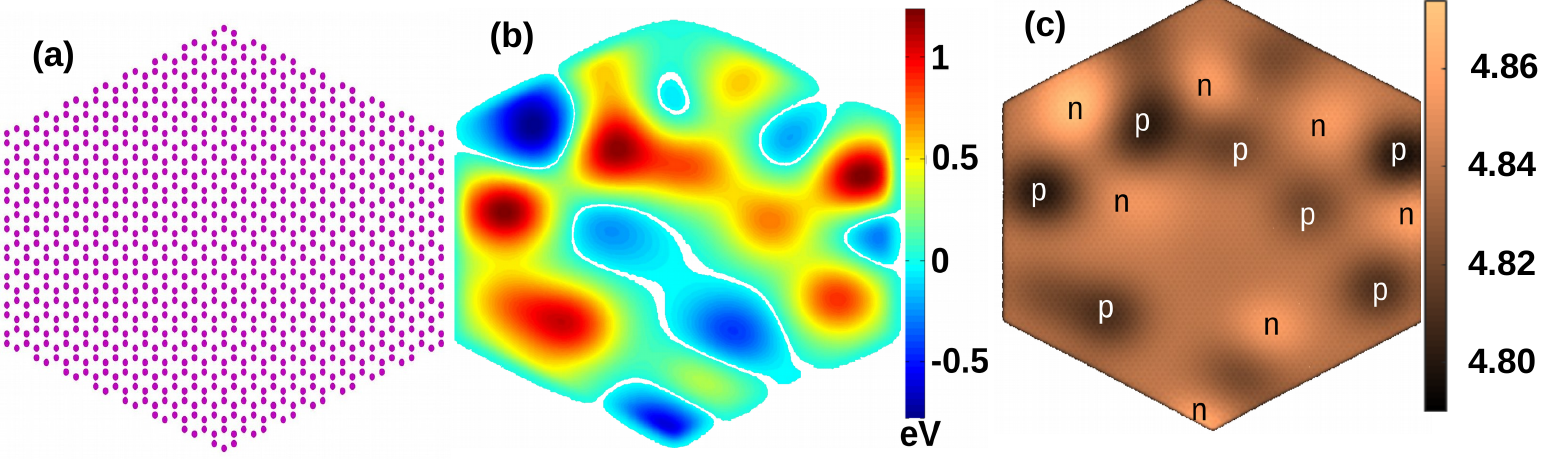}
\caption{(Color online) (a) Lattice structure of hexagonal armchair
  edged GQD. (b) Impurity potential landscape. (c) Electron-hole puddle
  formation achieved by MFH model where local charge accumulation is
  indicated as n and p puddles. }
\end{figure}

In order to model the long-range disorder due to charge impurities
caused by substrate effects, we use a superposition of Gaussian
electrostatic potentials $V_{imp}$ which are determined randomly to
have a smooth potential landscape (see Fig. 1b) on the
GQD. Impurity potential is written as:
\begin{align} 
V_{imp}(r_i) = \sum_k V_k \ exp\Big[{-\frac{(\vec{r}_i-\vec{R}_k)^2}{2\sigma^2}}\Big]
\end{align}

where $V_k$ is chosen to be the potential peak value which is randomly
generated between $-V_{max}< V_{k} < V_{max}$ values for an impurity at $R_k$, characterizing the strength of the disorder. For
most of the calculations, we take $V_{max}=t_{nn}/3$ giving a medium disorder strength. However, the effect of strong ($V_{max}=t_{nn}$) and weak ($V_{max}=t_{nn}/5$) disorder is also investigated (see Fig. 5).  The width of the potential, $\sigma$, is
determined to be 10 times the lattice constant in order to simulate
long-range lattice scatterers\citep{puddle1}. For 5 nm (1014 atoms), 13 nm (5514 atoms) and 18 (10806 atoms) nm wide GQDs,
respectively 4, 20 and 40 source point of impurities are randomly created to have approximately similar source point densities (but different form of distribution of source points) for each GQD.
Moreover, we considered 5 different randomly chosen potential
configurations for each QD size. The main effect of long-range
disorder on the electronic densities is the formation of electron-hole
puddles\cite{puddle1,ulasmenafdevrim}, as seen from Fig. 1c, obtained
by subtraction of the positive background charge from MFH electronic
density. The effect of the electron-hole puddles on the optical
properties will be investigated below using TB, MFH and
CI approaches.

Interaction of GQD's electrons with photons are evaluated within
electric dipole approximation by the interaction Hamiltonian
$H_{int}=\mathbf{E}\cdot\mathbf{r}$ where $\mathbf{E}$ is the photon's
electric field and $\mathbf{r}$ is the electron's position. Hence, one
can obtain absorption spectrum by using light-matter interaction which
is described as:
\begin{gather}
A(\omega)=\sum_{f} \frac{4\pi^2\alpha E_{fi}{\mid <f\mid \mathbf{r} \mid i> \mid}^{2}\delta(\hbar\omega-E_{fi})}{Area}
\end{gather}
where $\alpha$ is fine structure constant, $Area$ is area of the QD, $E_{fi}$ is the difference between initial and final energies,  $<f\mid \mathbf{r} \mid i>$ denotes  dipole matrix element, $\mid
i>$ and $\mid f>$ denote initial and final occupied molecular orbitals
, respectively,
obtained by TB and MFH model.

On the other hand, we obtain absorption spectrum which includes many-body correlations as:
\begin{gather}
A(\omega)=\sum_{f}\frac{4\pi^2\alpha E_{fi}{\mid <f\mid \mathbf{P^{\dagger}} \mid gs> \mid}^{2}\delta(\hbar\omega-E_{fi})}{Area}
\end{gather}
where $\alpha$ is fine structure constant, $Area$ is area of the QD, $E_{fi}$ is the difference between initial (ground state) and final energies of exciton, $\mathbf{P^{\dagger}}$ annihilates a photon and adds an exciton
to the ground state of the GQD. The final excitonic state  $\mid f>$ is obtained from CI calculations, and $\mid gs>$ is the ground state.

\section{Results}

\begin{figure}
\includegraphics[scale=0.335]{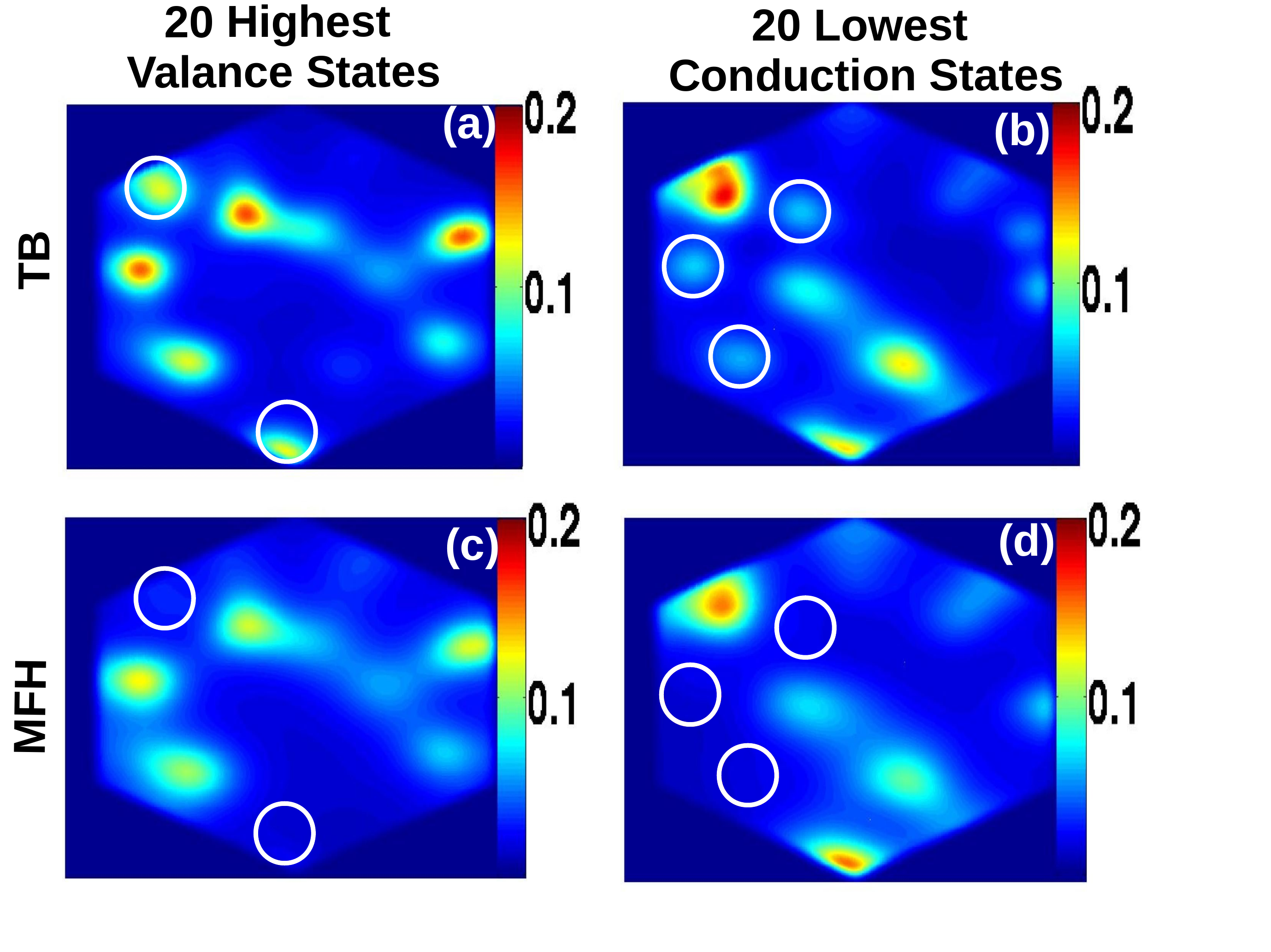}
\caption{(Color online) Electronic density corresponding to the 20
  highest valence states (left panels), and the 20 lowest conduction
  states (right panels), obtained from TB (upper panels) and MFH (lower
  panels) model of the structure 18 nm width size. Electron-electron
  interactions wash out abnormal localized states near Fermi level as
  indicated by white circles.}
\end{figure}

In Fig.2, we investigate electronic densities corresponding to 20
lowest conduction and 20 highest valence states obtained from TB and
MFH calculations for the largest GQD structure that we studied, which
has 10806 atoms giving a width of 18 nm (see corresponding potential landscape in Fig. 1b). We note that we repeated all
the calculations for 5 different random potential landscape (for each QD size) and
observed similar behaviors. In the TB results, in addition to valance
states accumulated around peaks and conduction states around troughs
(see Fig. 2a and 2b) as expected, we also observe abnormal valance states
around troughs and conduction states around peaks (shown in circles,
to be compared with Fig.1b). In fact, those abnormal states are an artifact of the TB method which is better suited for systems with homogeneous and neutral charge distributions. In our system, the charge density fluctuates strongly due to random disorder and the energy gap between valence and conduction states is not large enough to protect hole states from mixing with electron states. Thus, a mean-field correction to the TB method must be included. Indeed, when electron-electron interactions
are included through MFH calculations, electronic density fluctuations are reduced in almost all area of the QD and the abnormal localized states are washed out (see Fig. 2c and 2d). 
Similar behavior was also observed in graphene nanoribbons\cite{ulasmenafdevrim}.
As we will see, the rearrangement of electron-hole puddles through 
electronic interactions has an important effect on optical properties.


\begin{figure*}
\includegraphics[scale=0.48]{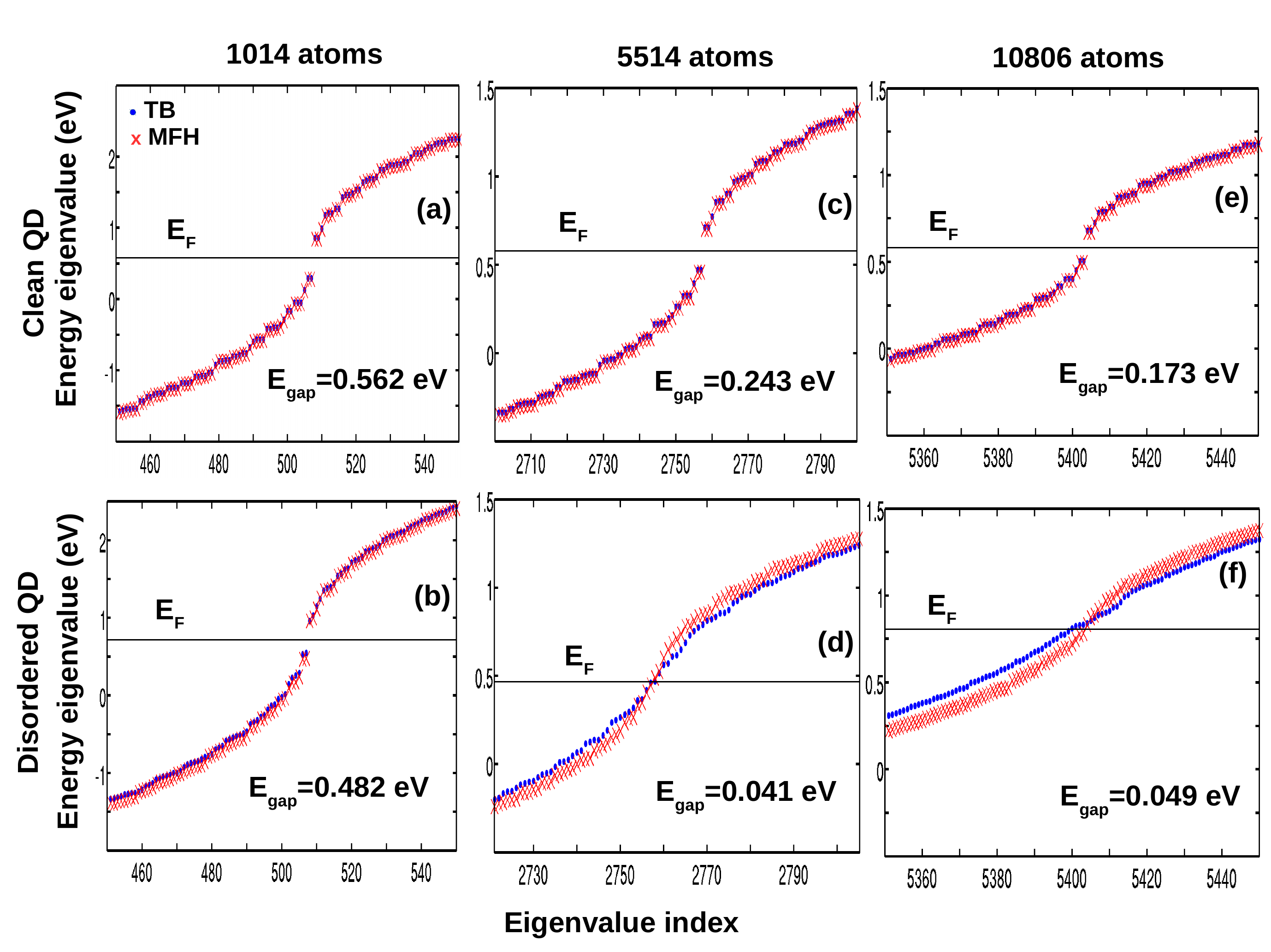}
\caption{(Color online) Energy spectra for clean (upper panels) and
  disordered (lower panels) GQDs obtained by TB and MFH. Fermi energy
  level $E_{F}$ is determined to be in the mid-point between valance
  and conduction band.}
\end{figure*}

Energy spectra of clean (upper panels) and disordered (lower panels)
GQDs having width size of 5 nm (1014 atoms), 13 nm (5514 atoms) and 18
nm (10806 atoms) obtained by TB and MFH model are shown in Fig.3. For
each case, the energy gap $E_{gap}$ between between lowest unoccupied
conduction state and highest occupied valence state obtained from the
MFH calculations is indicated as well.  As expected, $E_{gap}$
decreases more rapidly as a function of size when impurities are
present. More interestingly however, for larger size disordered GQDs
the difference between TB and MFH spectra become pronounced indicating
that when charge inhomogeneities (due to electron-puddle formation)
are present it is important to include the effects of electronic
interactions. Similar behavior was also observed for other random
potential configurations that we have tested.

\begin{figure*}
\includegraphics[scale=0.48]{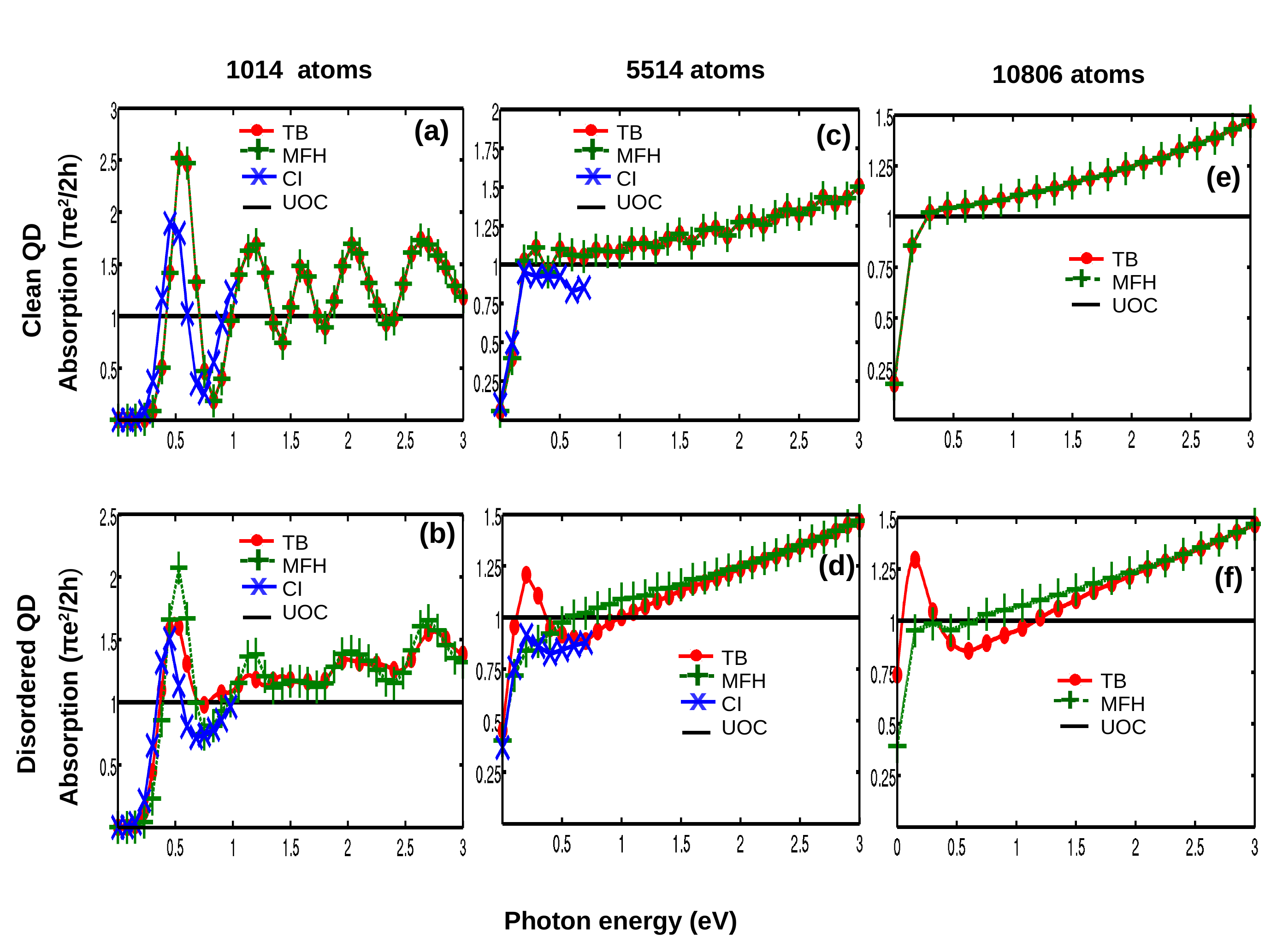}
\caption{(Color online) Absorption Spectrum for clean (upper panels)
  and disordered (lower panels) GQDs obtained by TB, MFH and excitonic
  effect with CI model. UOC is indicated by black line. In clean GQDs, as the size of GQDs increase, a plateau develops near the UOC at low energies, before a sudden drop occurs due to finite size effects.}
\end{figure*}

Figure 4 shows absorption spectra curves corresponding to the GQDs considered
in Fig.3 for energies up to 3 eV. The absorption spectra are calculated using equations (4) and (5) with a Gaussian broadening (0.1 eV) of delta functions in order to obtain continuous curves, within TB
(line doted curve, red color online), MFH (line plus signed curve, green color
online) and CI (line cross signed curve, blue color online) approaches.
The UOC is indicated by black
line as a reference. For clean GQDs, there is no noticeable difference
between the TB and MFH results, consistent with the results in
Fig.3. We note that, as the system size increases, absorption curves approach the UOC value at low energies, until a sudden drop occurs due to finite size effects. For
the CI calculations, 100 highest valence and 100 lowest conduction
states were included to form a many-body basis set of 10000 excitonic
states, to ensure convergence for energies up to 0.75 eV. As seen from
Fig.4a and Fig.4b, the main effect of excitonic correlations is to
red shift the absorption spectrum\cite{electronicoptic1} followed by a slight decrease in the peak value.  For
GQDs larger than 13 nm (5514 atoms), it was not possible to calculate the CI absorption spectrum
due to computational limits.

When disorder is present, we observe a dramatic difference between the
TB and MFH results, shown in Fig.4b,d,f. This is mainly due to the
redistribution of electron-hole puddles discussed in Fig.2. For the
medium and large size GQDs without electronic interactions, in TB
calculations, both electrons and hole puddles may be present at the
same locations, giving rise to stronger electric dipole coupling, thus
higher absorption values in average at lower energies. Note that the
situation is different for the GQD with 1014 atoms, since the puddle
formation is much less well defined as the size of the QD is reduced, 
and the specific form of the disorder landscape has a bigger role. For medium size GQD, however, a disorder peak reappears at low energies when excitonic correlations are taken into account. This is due to the fact that excitonic interactions rearranges the electron and hole distributions within the disorder troughs and peaks, as we discuss below in Fig.6. 
We note that the CI results obtained for the disordered GQD with
5514 atoms is consistent with the experimental results for graphene sheet
\cite{optical1,opticconductance1,opticconductance2}. 

\begin{figure}
\includegraphics[width=8.6cm,height=13cm]{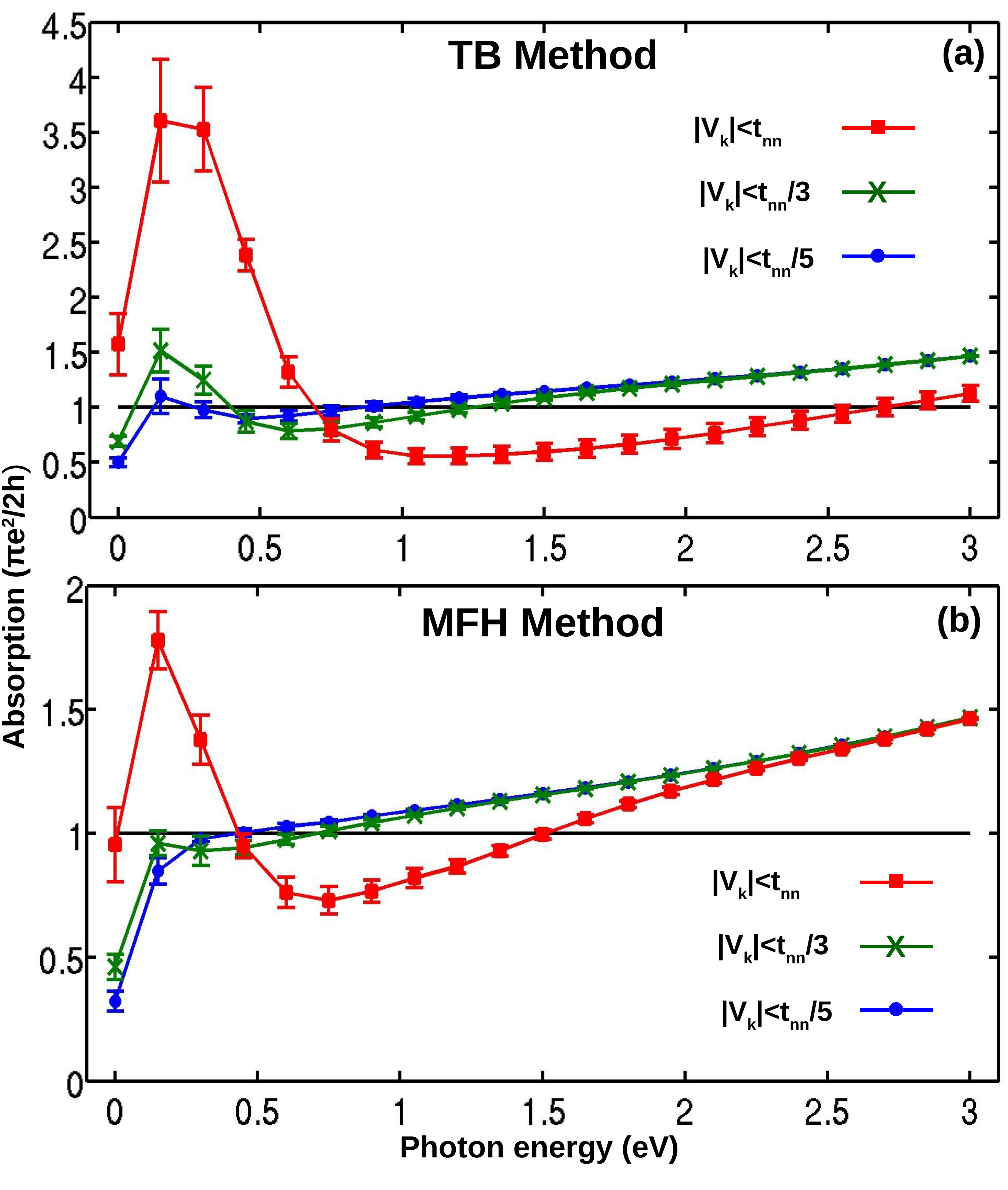}
\caption{(Color online) Average absorption spectrum curves (each curve corresponds to average of 5 different configurations) for three different impurity potential peaks  obtained by TB (upper panel) and MFH (lower panel) method with estimated error bars for the structure containing 10806 atoms. UOC is indicated by black line.}
\end{figure}

To see effects of various impurity potential strength on absorption spectrum obtained by TB and MFH methods (see Fig. 5a-b), we compare spectrum curves (each spectrum curve corresponds to average of five different samples shown with errorbars having width of twice the standard error) containing three different impurity potential strength peak values of $t_{nn}$ (line squared curve, red color online), $t_{nn}/3$ (line cross signed curve, green color online) and $t_{nn}/5$ (line doted curve, blue color online), for the largest QD structure. For the strong impurity potential strength ($|V_{k}| < t_{nn}$), both TB and MFH results deviate significantly from UOC line indicating that the system is in a strongly non-perturbative regime, and meanfield electron interactions are not sufficiently strong to wash out the impurity peak. However, for medium potential strength ($|V_{k}| < t_{nn}/3$) and small potential strength $|V_{k}| < t_{nn}/5$, the low energy absorption obtained from MFH remains always below the UOC line within our error bars.

\begin{figure}
\includegraphics[scale=0.43]{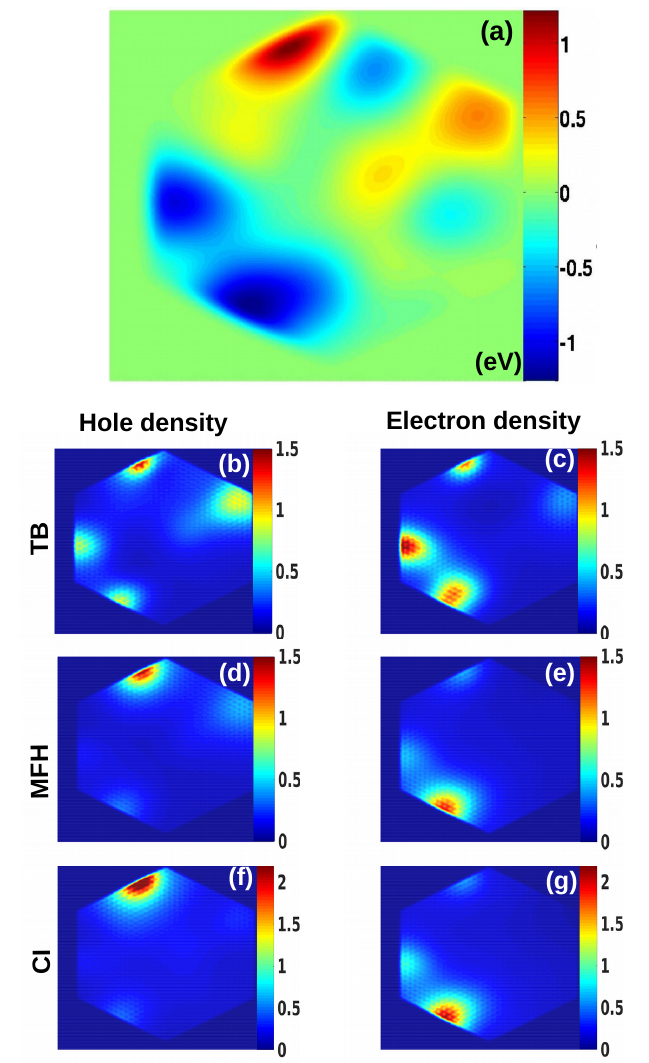}
\caption{(Color online) (a) Impurity potential for the
  structure containing 5514 atoms. (b-g) show corresponding electron and hole densities weighted with absorption probabilities in the energy range between 0 eV and 0.3 eV obtained by TB, MFH and CI
models, respectively. }
\end{figure}

In order to investigate the effect of excitonic correlations further, in Fig.6 we plot the electron and hole densities weighted with absorption probabilities 
in the energy range between 0 eV and 0.3 eV for the 5514
atom GQD, obtained from TB , MFH and CI calculations. As discussed earlier, mean-field interactions smooth the puddles so that excitonic
hole states are now localized only on peaks, and the electron states
are localized on troughs as seen in Fig.6d-e. On the other hand, the correlations have
a less dramatic effect on the density distribution, but the electron
states are now slightly more localized on a potential trough that is
closer to the hole puddle (see Fig.6f-g). Indeed, the electron-hole attraction is
favoured in the CI calculations minimizing the average distance between the
electron and the hole, thus increasing the electric dipole strength and
the absorption at lower energies.

\section{Conclusions}
In conclusion, we have investigated electronic and optic properties of
three different sizes of clean and disordered hexagonal armchair-edged
GQDs by applying tight-binding, mean-field Hubbard and configuration
interaction models. Long-ranged disorder give rise to formation of
electron-hole puddles, which are, however poorly described by the tight-binding
model alone. Electronic interactions in the mean-field picture reorganize
the electron-hole puddles, strongly affecting the dipole moments
between the low-energy states in the electronic spectrum. Hence,
inclusion of electronic interactions are found to be important in 
order to correctly describe the optical properties. As the system size is
increased to 18 nm, absorption spectra obtained from
configuration interaction method approach the experimental results leading
to observation of universal optical conductivity\cite{opticconductance1,opticconductance2}.


\end{document}